\documentclass[twoside,12pt]{article}
\usepackage{epsfig}

\def\Journal#1#2#3#4{{#1} {#2} (#4) #3 }
\def\Adva{{\em Adv. Nucl. Phys.}}
\def\EPJA{{\em Eur. Phys. J.} A}
\def\PPNP{{\em Prog. Part. Nucl. Phys.}}

\def\NPA{{\em Nucl. Phys.} A}

\def\NPB{{\em Nucl. Phys.} B}

\def\PLB{{\em Phys. Lett.} B}

\def\PRL{\em Phys. Rev. Lett.}

\def\PREP{\em Phys. Rep.}

\def\PRD{{\em Phys. Rev.} D}
\def\PRC{{\em Phys. Rev.} C}

\def\ZPA{{\em Z. Phys.} A}

\def\Re{{\rm Re}}
\def\Im{{\rm Im}}
\newcommand{\be}{\begin{equation}}
\newcommand{\ee}{\end{equation}}
\newcommand{\bea}{\begin{eqnarray}}
\newcommand{\eea}{\end{eqnarray}}

\begin{document}

\title{ \vspace{1cm} Finite-width QCD sum rules for 
$\rho$ and $\omega$ mesons\footnote{Dedicated to the memory of O.P. Pavlenko}}

\author{B.\ K\"ampfer, S.\ Zschocke\\ \\ 
Forschungszentrum Rossendorf, PF 510119, 01314 Dresden, Germany}

\maketitle

\begin{abstract} 
We present a combined analysis of the in-medium
behavior of $\rho$ and $\omega$ mesons 
within the Borel QCD sum rule taking into account finite widths.
\end{abstract}

%\eject

%\tableofcontents

\section{Introduction}

The experiments at the High-Acceptance Di-Electron Spectrometer HADES \cite{HADES}
are aimed at verifying predictions of the behavior of light vector mesons 
in nuclear matter. Due to the decay channel $V \to e^+ e^-$ a measurement
of the escaping di-electrons can reveal directly the properties of the parent
mesons, $V = \rho, \omega, \cdots$, since the interaction probability of the
$e^\pm$ with the ambient strongly interacting medium is small.   

In relativistic heavy-ion collisions temperature effects play an important role,
which also cause a change of the properties of vector mesons. Indeed, the experiments
of the CERES collaboration at the CERN-SPS \cite{CERES} can only be explained
by assuming strong medium effects, in particular for the $\rho$ meson
(cf.~\cite{Rapp_Wambach,Gallmeister} and further references therein). This seems to be
confirmed at higher beam energies, 
as delivered by the Relativistic Heavy Ion Collider, since the $\rho$ meson,
as measured via the $\pi^+ \pi^-$ decay channel, suffers some modification \cite{RHIC}.
In contrast, in heavy-ion collisions at typical SIS18 energies, i.e., at beam
energies around 1 AGeV, the in-medium behavior of vector mesons in compressed
nuclear matter can be studied, where temperature effects are small and may be neglected.
Complementary to heavy-ion collisions one can seek for in-medium modifications
in reactions of hadronic projectiles \cite{KEK}
or real and virtual photons \cite{Metag} at nuclei, as already
at nuclear saturation density sizeable modifications of vector mesons
are predicted.

There exists a vastly extended literature on the in-medium modification of
hadrons. We mention here only the Brown-Rho scaling hypothesis, according to which
a mass shift of a vector meson is directly interrelated to a change of the chiral
condensate \cite{BR_scaling}, and the vector manifestation \cite{Harada_Yamawaki},
the effective Lagrangian approach \cite{Klingl_Weise}, purely hadronic approaches
\cite{Mosel1,Lutz}, and QCD sum rules \cite{Hatsuda,Ko,Mosel2,Leupold}. 
QCD sum rules \cite{pioneers} follow the idea of duality (cf.~\cite{Cohen})
by relating quantities expressed by partonic (quark and gluon) degrees of freedom
with hadronic observables. We take here the attitude to assume that the
partonic quantities are given and examine the QCD sum rule to elucidate
the in-medium change of the $\rho$ and $\omega$ meson on a common footing.
The corresponding current operators, expressed by the interpolating quark 
field operators $u$ and $d$ have the form 
$J_\mu^\rho   = \frac12 ( \bar u \gamma_\mu u - \bar d \gamma_{\mu} d)$ and
$J_\mu^\omega = \frac12 ( \bar u \gamma_\mu u + \bar d \gamma_{\mu} d)$,
suggesting that the same partonic quantities enter the sum rules.
A common treatment of $\rho$ and $\omega$ mesons is necessary since our
schematic transport model studies \cite{Gyuri} show that the relative shift
of the peak positions and in-medium broadenings must be known to arrive at
firm predictions concerning the chances to identify both mesons in
experiments performed at HADES. 

In a broader context such studies address the phenomenon ''mass''. Within the
standard model, masses of quarks and leptons are generated by spontaneous 
symmetry breaking (in the Higgs mode) in the electro-weak sector, while the
spontaneous symmetry breaking (in the Goldstone mode) explains the features
of the hadronic mass spectrum: light Goldstone bosons ($\pi, K, \eta$ with
finite masses generated by an additional explicit symmetry breaking)
and heavier hadronic states emerge \cite{Weise_suvey}. Within the QCD sum
rule approach \cite{pioneers,Cohen} the hadron masses can be related to
condensates which represent non-perturbative features of the QCD vacuum.
This is highlighted by the Gell-Mann--Oakes--Renner relation,
$m_\pi^2 f_\pi^2 = - 2 m_q \langle \bar q q \rangle$ (which is actually related
to PCAC), and the Ioffe formula, $m_N = - 8 \pi^2 \langle \bar q q \rangle / M^2$
(cf.\ \cite{Weise_Buch}).
Both expressions, which are in leading order, suggest that hadron masses
are tightly related to the chiral condensate $\langle \bar q q \rangle$,
which in turn is a measure of the order parameter of the chiral symmetry
breaking.

Since $\langle \bar q q \rangle$ changes with changing temperature \cite{Karsch}
and density (cf.\ Fig.~1 in \cite{Zschocke0}), one can expect an simultaneous
change of the hadron masses. This was the very idea of the Brown-Rho scaling
hypothesis \cite{BR_scaling} which is modified according to \cite{Hatsuda}.
This promoted very much the physics programme at HADES with the motivation
to verify, via an observed change of the hadron masses, 
the chiral condensate's change.

To call a loosely spoken analogy, one is seeking for the QCD analog of the
Zeeman/Stark effects: an external field (here: strong interaction mediated
by surrounding hadrons) changes the excitation spectrum of an atom
(here: the hadronic excitation spectrum above the QCD ground state).
Very often the notion ''mass shift'' is used as short hand notation
for a shift of the peak in the spectral function.

Our paper addresses such mass shifts of $\rho$ and
$\omega$ mesons on the basis of the Borel QCD sum rule;  
it is organized as follows: In section II we recapitulate the 
basics of the QCD sum rule approach. In section III we present
a combined study of the in-medium behavior of $\rho$ and $\omega$
mesons with accounting for effects of finite widths.
The $\rho - \omega$ mass splitting is discussed in section IV.
The conclusions can be found in section V.
  
\section{QCD sum rule}

Within QCD sum rules (QSR) the in-medium vector mesons $V=\rho, \omega$ are 
considered as resonances  
in the current-current correlation function
\be
\Pi_{\mu \nu} (q , n) = i \int d^4 x \;{\rm e}^{i q \cdot x} 
\langle {\cal T} \; J_\mu^V (x)\; J_\nu^V (0)\rangle_n\;,
\label{eq_5}
\ee
where $q_{\mu}=(E, {\bf q})$ is the meson four momentum, ${\cal T}$ denotes the 
time ordered product of the respective meson current operators 
$J_\mu^V (x)$,
and $\langle \cdots \rangle_n$ stands for the expectation value in medium. 
In what follows, we focus on the ground state of low-density baryon matter approximated 
by a Fermi gas with nucleon density $n$. 
We consider isospin symmetric nuclear matter, where the $\rho - \omega$ 
mixing effect is negligible. 

The correlator (\ref{eq_5}) can be reduced to  
$\frac{1}{3} \Pi_{\mu}^{\mu} (q^2, n) = %\sum\limits_{V=\rho,\omega} 
\Pi^{(V)} (q^2, n)$ for a vector meson at rest, ${\bf q}=0$,  
in the rest frame of matter. 
In each of the vector meson channels the 
corresponding correlator 
$\Pi^{(V)}(q^2, n)$ satisfies the twice subtracted dispersion relation, 
which can be written with $Q^2 \equiv -q^2 = -E^2$ as 
\be
\frac{\Pi^{(V)} (Q^2)}{Q^2} = \frac{\Pi^{(V)} (0,n)}{Q^2} - \Pi^{(V) '} 
(0) - Q^2 
\int\limits_0^{\infty} \; ds \frac{R^{(V)}(s)}{s (s + Q^2)}\;,
\label{eq_10}
\ee
with $\Pi^{(V)} (0,n) = \Pi^{(V)} (q^2=0, n)$ and $\Pi^{(V) '} (0)=
\frac{{\rm d} \Pi^{(V)} (q^2)}{{\rm d} q^2}|_{q^2=0}$ 
as subtraction constants, and 
$R^{(V)} (s) \equiv - \Im \Pi^{(V)} (s, n) /(\pi s)$.

For large values of $Q^2$ one can evaluate 
the r.h.s.\ of eq.~(\ref{eq_5}) 
by the operator product expansion (OPE) leading to
\be
\frac{\Pi^{(V)}(Q^2)}{Q^2} = - c_0\; {\rm ln}(Q^2) + \sum\limits_{i=1}^{\infty}
\;\frac{c_i}{Q^{2i}}\;,
\label{eq_15}
\ee
where the coefficients $c_i$ include the Wilson 
coefficients and the expectation values of the corresponding products of the 
quark and gluon field operators, i.e. condensates.
Performing a Borel transformation  
of the dispersion relation (\ref{eq_10}) with appropriate 
parameter $M^2$ and taking into account the 
OPE (\ref{eq_15}) one gets the basic QSR equation 
\be
\Pi^{(V)} (0,n) + \int\limits_0^{\infty} d s \,R^{(V)} (s)\, {\rm e}^{-s/M^2} = 
c_0 M^2 + \sum\limits_{i=1}^{\infty} \frac{c_i}{(i-1)! M^{2 (i-1)}}\,.
\label{eq_20}
\ee
The advantage of the Borel transformation is 
(i) the exponential suppression of the high-energy part of $R^V (s)$, and 
(ii) the possibility to suppress higher-order
contributions to the r.h.s.\ sum. Choosing sufficiently large values
of the internal technical parameter $M$ one can truncate the sum
in a controlled way, in practice at $i = 3$. 

For the subtraction constants $\Pi^{(V)} (0,n)$ in eq.~(\ref{eq_10}) 
we use $\Pi^{(\rho)} (0,n) = n/(4 M_N)$, 
$\Pi^{(\omega)} (0,n) = 9 n/(4 M_N)$, 
which are actually the Thomson limit of the 
$VN$ scattering processes, but also coincide with Landau damping terms elaborated 
in \cite{Hofmann} for the hadronic spectral function entering the dispersion 
relation without subtractions. 
For details about the connection of subtraction constants 
and Landau damping term we refer the interested reader to \cite{WW}.

In calculating the density dependence of the 
condensates entering the coefficients $c_i$ we employ the standard 
linear density approximation, which is valid for not too large density. 
This gives for the chiral quark condensate 
$\langle \overline{q} q\rangle_n = \langle \overline{q} q\rangle_0 
+ \frac{\sigma_N}{2 m_q} n\;$,
where we assume here isospin symmetry for the light quarks, 
i.e. $m_q = 5.5$ MeV and 
$\langle \bar q q \rangle_0 = - (0.24 \, {\rm GeV})^3$.
The nucleon sigma term is $\sigma_N = 45 $ MeV.
The gluon condensate is obtained as usual employing the QCD trace anomaly
$\langle\frac{\alpha_s}{\pi} {\rm G^2}\rangle_n = 
\langle\frac{\alpha_s}{\pi} {\rm G^2}\rangle_0 - \frac{8}{9} M_N^0 \;n \;,$
where $\alpha_s=0.38$ is the QCD coupling constant and $M_N^0=770 $ MeV 
is the nucleon mass in the 
chiral limit. The vacuum gluon condensate is 
$\langle \frac{\alpha_s}{\pi} G^2\rangle_0 = (0.33\,{\rm GeV})^4$.

The coefficient $c_3$ in eq.~(\ref{eq_20}) contains also the  
mass dimension-6 4-quark condensates
(cf.\ \cite{Faessler} for a recent calculation of corresponding
matrix elements)  
$\langle (\bar q \gamma_{\mu}\lambda^{a} q)^2 \rangle_n$,
$\langle (\bar u \gamma_{\mu}\lambda^{a} u)(\bar d \gamma^{\mu}\lambda^{a} d)\rangle_n$,
$\langle (\bar q \gamma_{\mu}\lambda^{a} q)$
$(\bar s \gamma^{\mu}\lambda^{a} s)\rangle_n$,
and 
$\langle(\overline{q}\gamma_{\mu}\gamma^5 \lambda^{a} q)^2\rangle_n$
which are common for $\rho$ and $\omega$ mesons.
On this level, $\rho$ and $\omega$ mesons differ only by the condensate
$\pm 2 \langle (\bar u \gamma_\mu \gamma_5 \lambda^a u)
(\bar d \gamma^\mu \gamma_5 \lambda^a d)\rangle_n$ (cf.\ \cite{Leupold}),
causing the small $\rho - \omega$ mass splitting in vacuum \cite{pioneers}.
Keeping in mind the important role of the 4-quark condensate \cite{Zschocke1,Zschocke2}
for the in-medium modifications of the vector mesons, 
we employ the following parameterization
\be
\langle(\bar q \gamma_\mu \gamma^5 \lambda^a q)^2 \rangle_n =
\frac{16}{9} \langle\overline{q} q\rangle_0^2 \; \hat \kappa_0 \;
\left[1+\frac{\hat \kappa_N}{\hat \kappa_0}\frac{\sigma_N}{m_q 
\langle \bar q q \rangle_0}\;n\right]\;.
\label{eq_35}
\ee 
The parameter $\hat \kappa_0$ 
reflects a deviation from the vacuum saturation assumption. (The case
$\hat \kappa_0=1$ corresponds obviously to the exact vacuum saturation \cite{Cohen}.)
To control the deviation of the in-medium 4-quark condensate 
from the mean-field approximation we introduce the parameter $\hat \kappa_N$. 
An analog procedure applies for the other 4-quark condensates,
each with its own $\hat \kappa_0$ and $\hat \kappa_N$,
which sum up to a parameter $\kappa_0$ and a parameter $\kappa_N$.
As seen in eq.~(\ref{eq_35}) and eq.~(\ref{eq_40}) below,
$\kappa_N$ parameterizes the density dependence of the summed 4-quark condensates;
$\kappa_0$ is adjusted to the vacuum masses. 
Below we vary the poorly constrained parameter $\kappa_N$ to estimate 
the contribution of the 4-quark 
condensates to the QSR with respect to the main trends of the in-medium 
modification of the vector meson spectral function.
(Strictly speaking, $\kappa_0$ and
$\kappa_N$ differ for $\rho$ and $\omega$ mesons due to contributions 
of the above mentioned flavor-mixing condensate; in addition, in medium a twist-4
condensate makes further $\rho$ and $\omega$ to differ 
\cite{Mosel2}.
However, the differences can be estimated to be sub-dominant. Therefore, we use
in the present work one parameter $\kappa_N$, keeping in mind that it may slightly
differ for different light vector mesons.)

Using the above condensates and usual Wilson coefficients one gets as relevant terms
for mass dimension $\le 6$ and twist $\le 2$ \cite{Zschocke1,Zschocke2} 
\bea
c_0 &=& \frac{1}{8 \pi^2} \left(1 + \frac{\alpha_s}{\pi}\right)\;, \label{eq_2.6}\\
c_1 &=& - \frac{3 m_q^2}{4 \pi^2} \;,\\
c_2 &=& m_q \langle\overline{q} q\rangle_0 + \frac{\sigma_N}{2} \;n + 
\frac{1}{24} \left[\langle\frac{\alpha_s}{\pi} G^2 \rangle_0 
- \frac{8}{9} M_N^0 \;n\right]
+ \frac{1}{4} A_2 M_N \;n\;, \\
c_3 &=& - \frac{112}{81} \pi \;\alpha_s \;\kappa_0\; 
\langle\overline{q} q\rangle_0^{2} 
\left[1+\frac{\kappa_N}{\kappa_0}\frac{\sigma_N}{m_q 
\langle \overline{q} q\rangle_0}\;n\right]  
- \frac{5}{12} A_4 M_N^3 \;n \nonumber \\
& & + \frac{4 \alpha_s}{81 \pi f_\pi^2} \langle \bar q q \rangle_0^2 Q_0^2
(2 \pm 9) \left[ 1 + 
\frac{\sigma_N}{m_q \langle \overline{q} q\rangle_0} \; n \right].
\label{eq_40}
\eea
The terms with $A_{2,4}$ in $c_{2,3}$ correspond to the derivative condensates 
from non-scalar operators as a consequence of the breaking of Lorentz 
invariance in the medium. These condensates are proportional to the 
moments of quark and anti-quark distributions inside 
the nucleon at scale $\mu^2=1 {\rm GeV}^2$ 
(see for details \cite{Hatsuda}). Our choice of the 
moments $A_2$ and $A_4$ is $1.02$ and $0.12$, respectively.
The last line in eq.~(\ref{eq_40}) stems from the flavor mixing condensate,
$\langle \bar u \cdots u \bar d \cdots d \rangle_n$, which has been evaluated
with a technique similar to \cite{pioneers}.
$Q_0 \sim {\cal O} (200$ MeV) is a cut-off parameter from momentum
integrals. For our purposes we can neglect terms related to $Q_0$.

The value of $\kappa_0$ in eq.~(\ref{eq_40}) is related 
to such a choice of the chiral condensate $\langle\overline{q} q\rangle_0$ to 
adjust the vacuum vector meson masses. 
In our QSR we have used $\kappa_0=3$, obtaining 
$m_{\rho, \omega}(n=0)=777$ MeV close to the nominal
vacuum values. 
The ratio $\kappa_N/\kappa_0$ in the parameterization (\ref{eq_35}) is restricted by 
the condition 
$\langle(\overline{q} \gamma_{\mu} \lambda^a q)^2\rangle_n \le 0$,   
so that one gets 
$0\le \kappa_N \le 4$ as reasonable numerical limits when considering
$n \le n_0$, as dictated by our low-density approximation.

The case of finite baryon density \underline{and} temperature has been
considered in \cite{Zschocke1}.
Here we focus on density effects with the reasoning that temperature
effects below 100 MeV are negligible. 

To model the hadronic side of the QSR (\ref{eq_20}) we make the standard 
separation of the vector meson spectral density $R^{(V)}$ into resonance part 
and continuum contribution by means of the threshold parameter $s_V$
\be
R^{(V)}(s, n)= F_V \;\frac{S^{(V)} (s,n)}{s} \;\Theta(s_V-s) + 
c_0\; \Theta (s-s_V)\;,
\label{eq_45}
\ee
where $S^{(V)} (s,n)$ stands for the resonance peak in the spectral function;
the normalization $F_V$ is unimportant for the following consideration. 
In vacuum, this ansatz is justified since the time-reversed reaction,
$e^+ e^- \to V$, which is directly related to $R^{(V)}$, exhibits a prominent
vector meson peak at low energies and a smooth continuum at higher energies. 
The sum rule can be then cast into the form
\be
\frac{\int\limits_0^{s_V} ds \; S^{(V)} (s,n)\;{\rm e}^{-s/M^2}}
{\int\limits_0^{s_V} ds \; S^{(V)} (s,n) \,s^{-1} \; {\rm e}^{-s/M^2}} = 
\frac{\displaystyle c_0\,M^2\,[1-\left(1+\frac{s_V}{M^2}\right) {\rm e}^{-s_V/M^2}] 
- \frac{c_2}{M^2} - \frac{c_3}{M^4}}{\displaystyle c_0\,\left(1-{\rm e}^{-s_V/M^2}\right) 
+ \frac{c_1}{M^2} + \frac{c_2}{M^4} + \frac{c_3}{2 M^6} 
- \frac{\Pi^{(V)} (0,n)}{M^2}}.
\label{eq_70}
\ee

Given as such, the sum rule allows to test consistency of a particular model
for the spectral function $S^{(V)}$ as exercised, e.g., in \cite{Klingl_Weise}.
In the zero-width approximation, $S^{(V)} (s,n) \propto \delta (s - \tilde m_V^2 (n))$,
the l.h.s. of eq.~(\ref{eq_70}) becomes simply $\tilde m_V^2(n)$. In this sense one
could also consider the l.h.s. as averaged mass, denoted as $\bar m_V^2 (n)$,
keeping in mind that it represents a normalized moment of the hadron strength $S^{(V)}$.
Then the sum rule eq.~(\ref{eq_70}) determines the parameters $\tilde m_V^2$ or
$\bar m_V^2$ by the density dependence of the condensates, encoded in the   
coefficients $c_{1, 2, 3}$ from eqs.~(\ref{eq_2.6} $\cdots$ \ref{eq_40}),
and the subtraction constant $\Pi^{(V)} (0,n)$. Without further explication of
$S^{(V)}$ nothing can be deduced from eq.~(\ref{eq_70}) on in-medium mass shifts
and broadening (see, however, \cite{Hatsuda} where, after determining the
mass parameter $\tilde m_V (n)$ an estimate of the corresponding width is
attempted). 

\section{Finite width effects} %%%%%%%%%%%%%%%%%%%%%%%%%%%%%%%%%%%%%%

Our intention is now to evaluate the QCD sum rule eq.~(\ref{eq_70}) by taking into
account the finite widths of the vector mesons. Note that already in vacuum
the $\rho$ and $\omega$ widths differ noticeably, 150.7 MeV and 8.43 MeV,
respectively. This must be reflected in the form of $S^{(V)}$.
 
Ref.~\cite{Leupold} made for the $\rho$ meson the ansatz
\be
S^{(V)} (s,n) = \frac{\gamma_V (n) \Gamma_V (s)}
{(s -m_V^2 (n))^2 + (\gamma_V (n) \Gamma_V (s))^2}
\label{ansatz}
\ee 
with the two parameters, $m_V(n)$ and $\gamma_V(n)$,
and some parameterization of the width $\Gamma_V (s)$.
Clearly, the one  eq.~(\ref{eq_70}) cannot determine the two unknowns 
$m_V$ and $\gamma_V$,
rather only the correlation of $\gamma_V (m_V)$ is determined at a given density.
In line with the above ansatz (\ref{ansatz}), one can use a more 
realistic form for the resonance spectral density $S^{(V)}$ based on the 
general structure of the in-medium vector meson propagator 
\be
S^{(V)} (s, n) = - \frac{ \gamma_V (n) \, \Im \Sigma_V (s,n)}
{(s - {\stackrel{0}{m}}_V^2 (n) - \Re \Sigma_V (s,n))^2 + (\gamma_V (n) \, \Im \Sigma_V(s,n))^2} ,
\label{eq_50}
\ee
with $\Re \Sigma_V(s,n)$ and $\Im\ \Sigma_V(s,n)$ 
as real and imaginary parts of the in-medium vector meson 
self-energy. In the spirit of eq.~(\ref{ansatz}) \cite{Leupold} the meson 
mass parameter $m_V (n)$ and the width factor $\gamma_V (n)$ become 
density dependent in nuclear matter to have a degree of freedom for the
''request'' of the QSR.
This dependence is determined by 
the QCD sum rule eq.~(\ref{eq_70}) and mainly governed by the QCD condensates. 
(An analogous approach with $\gamma_\rho = 1$ was used in \cite{Ko}.) 
The in-medium vector meson mass is determined by the pole position of the 
meson propagator, i.e., 
$m_V^2 = {\stackrel{0}{m}}_V^2 (n) + \Re \Sigma_V (s = m_V^2 (n), n)$. 

Within the linear density approximation the vector meson self energy is 
given by 
\be
\Sigma_V (E,n) = \Sigma^{\rm vac}_V (E) - n\;T_{V N} (E)\;,
\label{eq_60}
\ee
where $E=\sqrt{s}$ is the meson energy, $\Sigma_V^{\rm vac} (E) 
= \Sigma_V (E,n=0)$, and $T_{V N} (E)$ is the (off-shell) forward 
meson-nucleon scattering amplitude in free space. 
The renormalized quantity  $\Sigma_\rho^{\rm vac}$ 
is summarized in the Appendix A in \cite{Zschocke3}.
For the $\omega$ meson we absorb as usual
$\Re \Sigma_\omega^{\rm vac}$ in ${\stackrel{0}{m}}_\omega^2$
and put simply 
$\Im \Sigma_\omega^{\rm vac} = - m_\omega \Gamma_\omega \Theta (E - 3 m_\pi)$
with the vacuum values of mass $m_\omega$ and width $\Gamma_\omega$.

The described framework is well defined, supposed $T_{VN}$ is reliably 
known.\footnote{Then, $\gamma_V = 1$ and $m_V (n) = m_V (0)$, and the QSR
acts merely as consistency check, as mentioned above.} 
Unfortunately, the determination of $T_{VN}$ is hampered by uncertainties
(cf.\ results in \cite{Klingl_Weise} and \cite{Lutz}). $\Im T_{VN}$ is more directly
accessible, while $\Re T_{VN}$ follows by a dispersion relation with
sometimes poorly known subtraction coefficients. Since our emphasis here is
to include the collision broadening and other finite width effects in the spectral
function, we absorb, as intermediate step, $\Re T_{VN}$ in 
${\stackrel{0}{m}}_V^2(n)$ thus neglecting a possible strong energy dependence. In such a way, the 
uncertainties of $\Re T_{VN}$ become milder since $m_V (n)$ is then mainly 
determined by the QSR. 
Neglecting the energy dependence of $\Re T_{VN}$ one discards a potentially rich structure
of $S^{(V)}$, such as, for instance, a double peak structure obtained
in \cite{Lutz} for the $\omega$ meson or in \cite{Mosel1} for the $\rho$ meson.
Afterwards the importance of $\Re T_{VN}$ is checked.

We take the needed scattering amplitude $T_{V N} (E)$ for $\rho$ 
and $\omega$ mesons from results of the detailed analysis of pion- 
and photon-nucleon scattering data performed recently in \cite{Lutz} 
on the footing of the Bethe-Salpeter equation approach with four-point 
meson-baryon contact interactions and a unitary condition for the coupled 
channels.

\begin{figure}[th]
\begin{center}
\begin{minipage}[t]{16 cm}
\epsfig{file=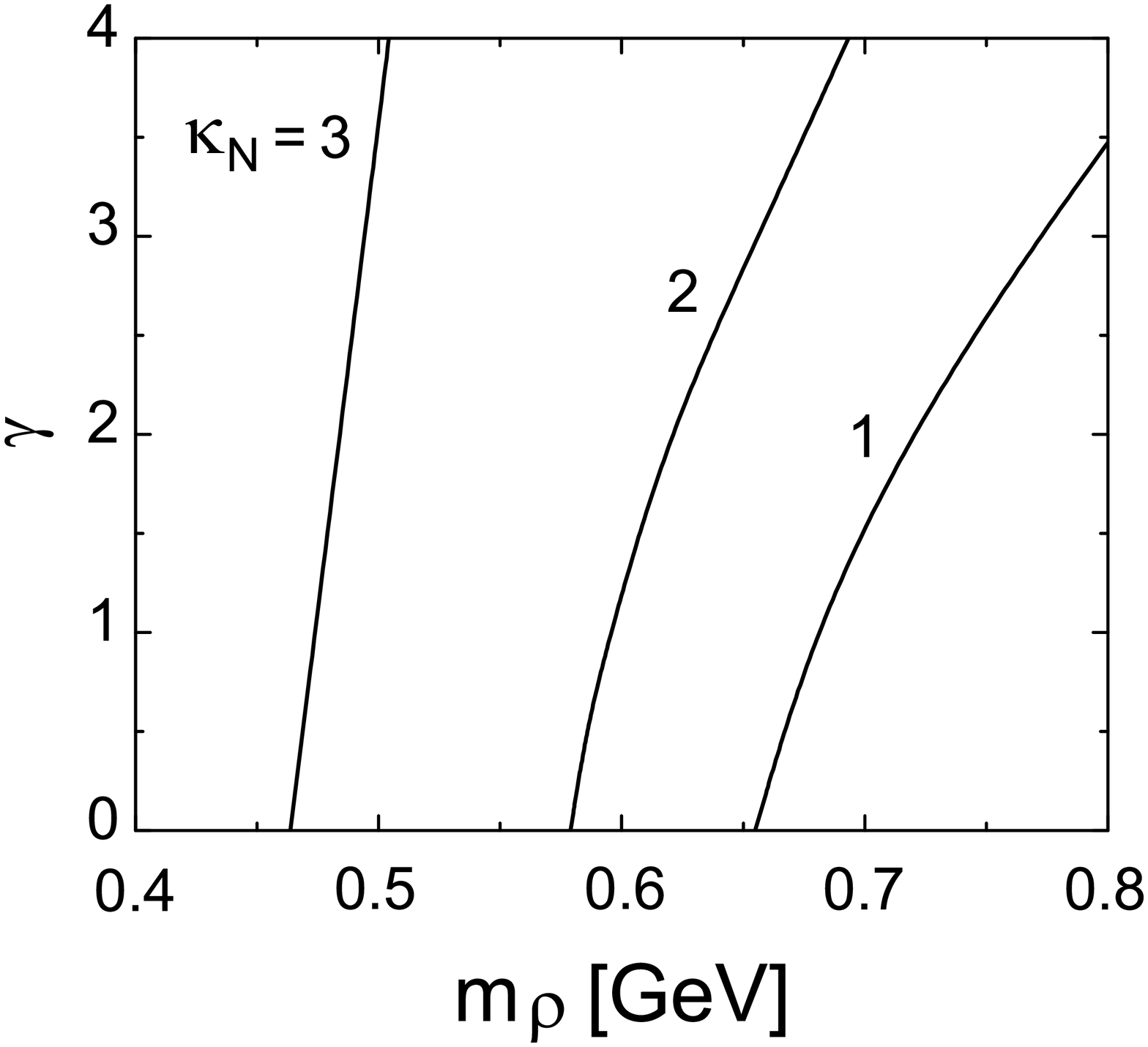,scale=0.33} \hfill
\epsfig{file=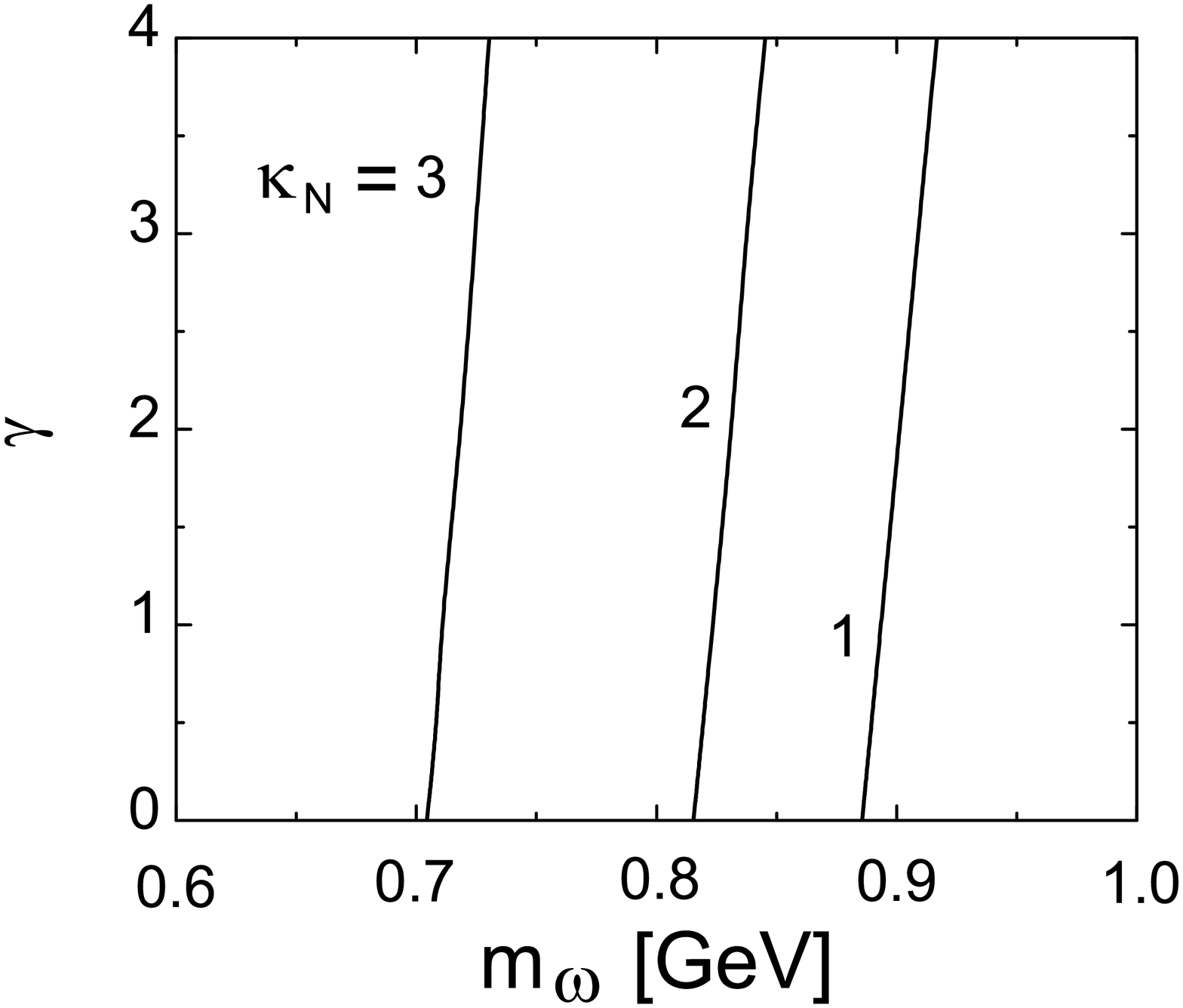,scale=0.33}
\end{minipage}
\begin{minipage}[t]{16.5 cm}
\caption{
Results of the QCD sum rule evaluations for $\rho$ (left panels)
and $\omega$ (right panels) at $n_0$. $T_{VN}$ is discarded.
\label{fig1}}
\end{minipage}
\end{center}
\end{figure}

\begin{figure}[th]
\begin{center}
\begin{minipage}[t]{16 cm}
\epsfig{file=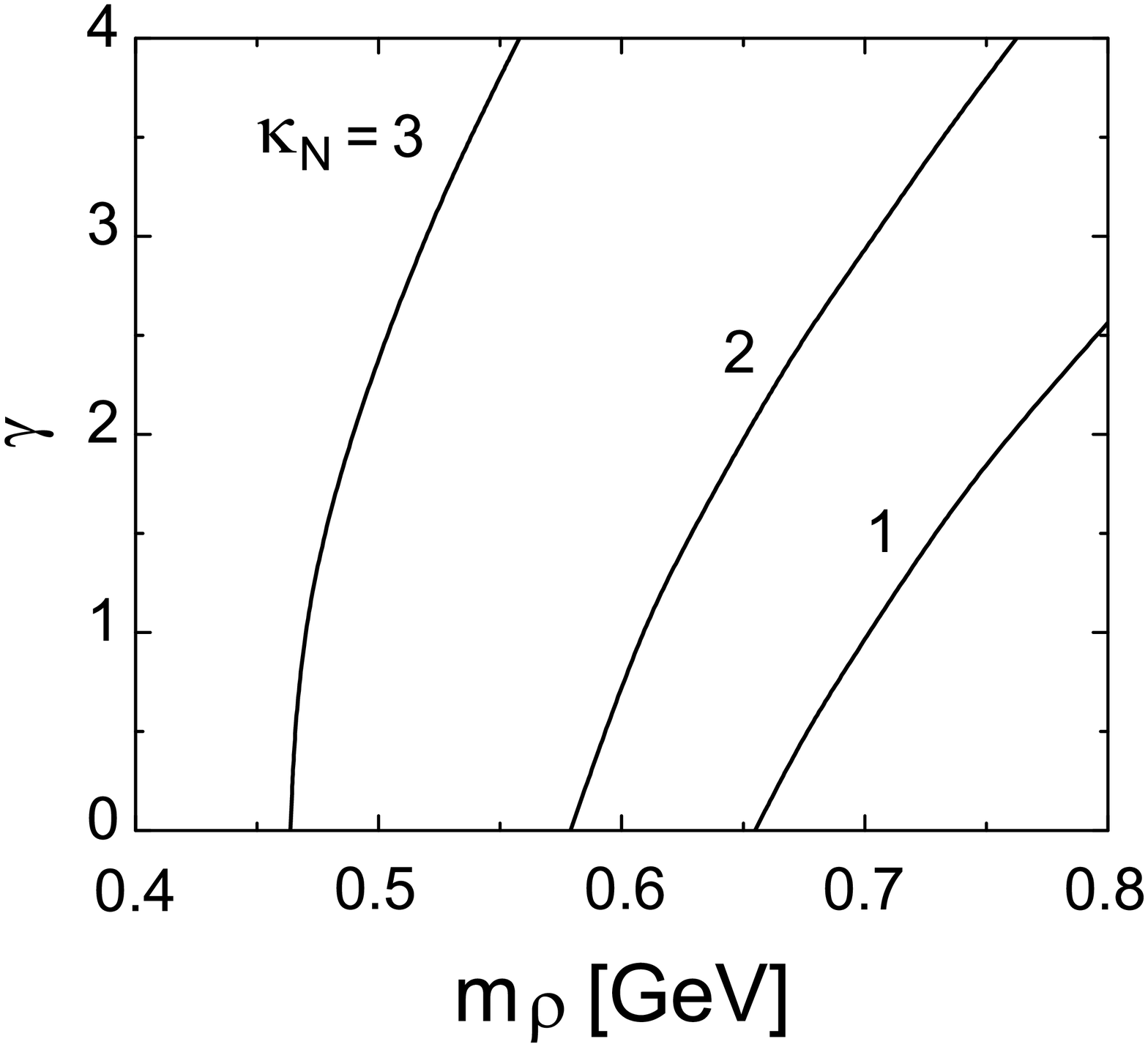,scale=0.33} \hfill
\epsfig{file=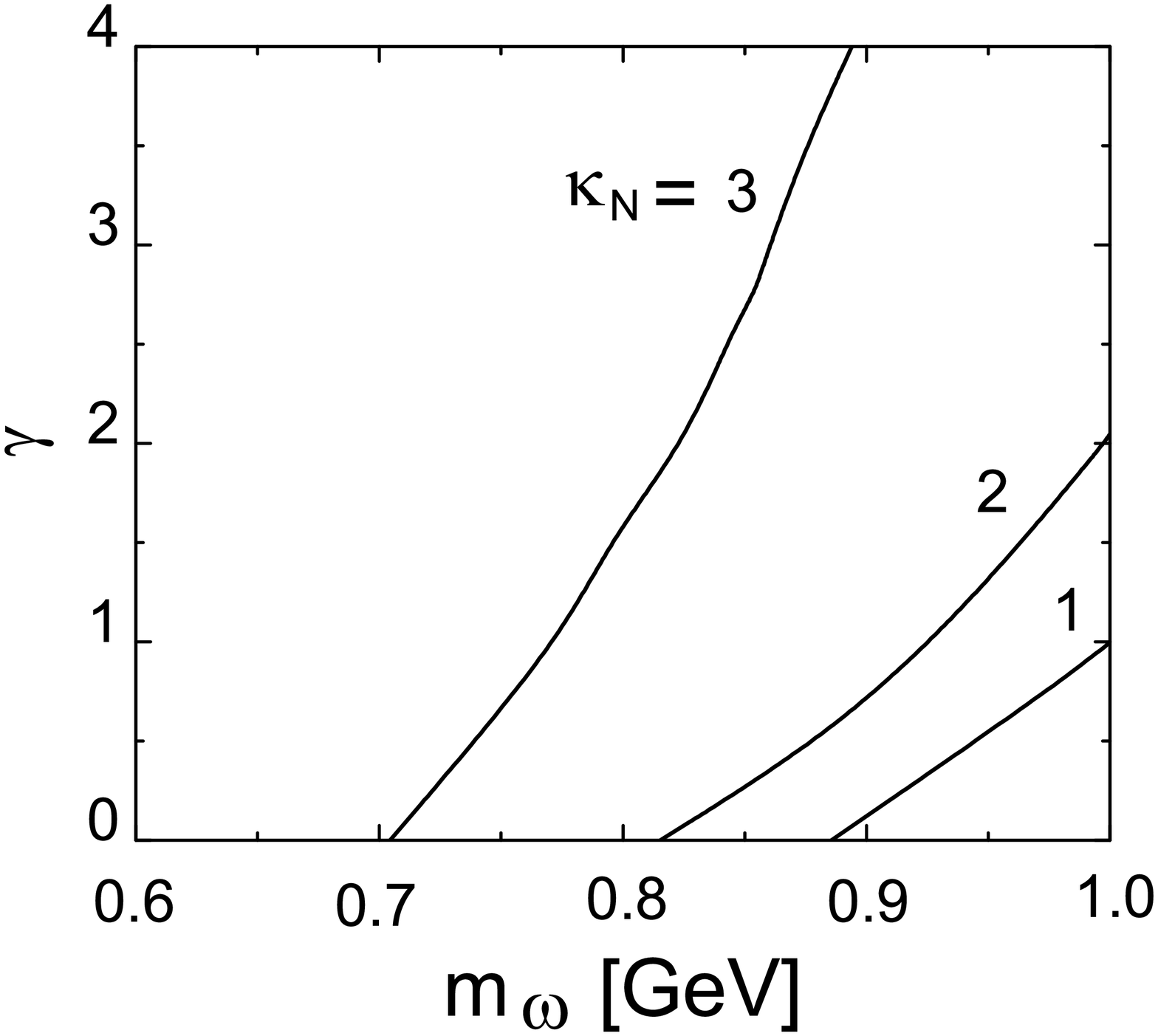,scale=0.33}
\end{minipage}
\begin{minipage}[t]{16.5 cm}
\caption{
As in Fig.~1 but with $\Im T_{VN}$ from \protect\cite{Lutz}.
\label{fig2}}
\end{minipage}
\end{center}
\end{figure}

\begin{figure}[tb]
\begin{center}
\begin{minipage}[t]{16 cm}
\epsfig{file=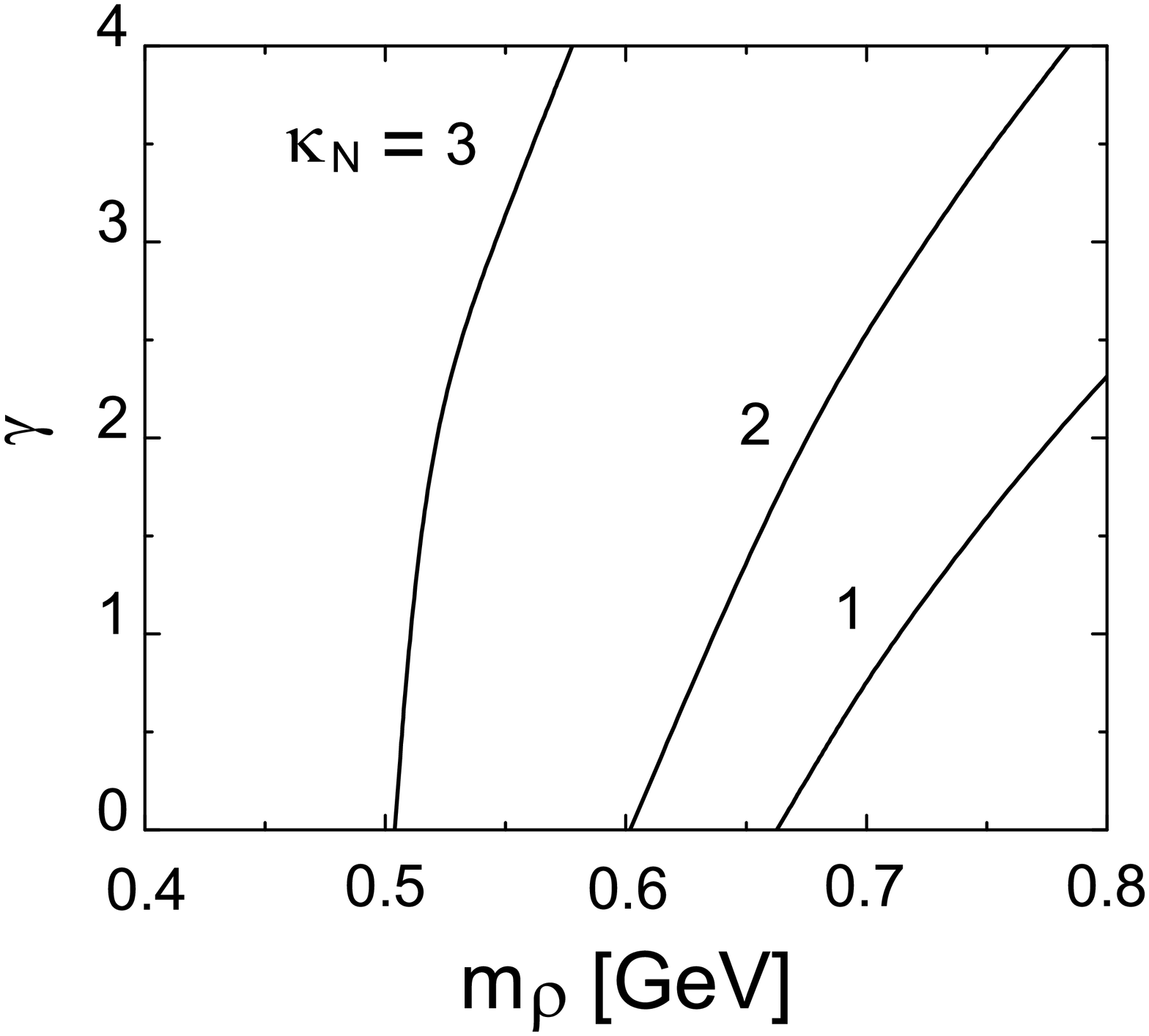,scale=0.33} \hfill
\epsfig{file=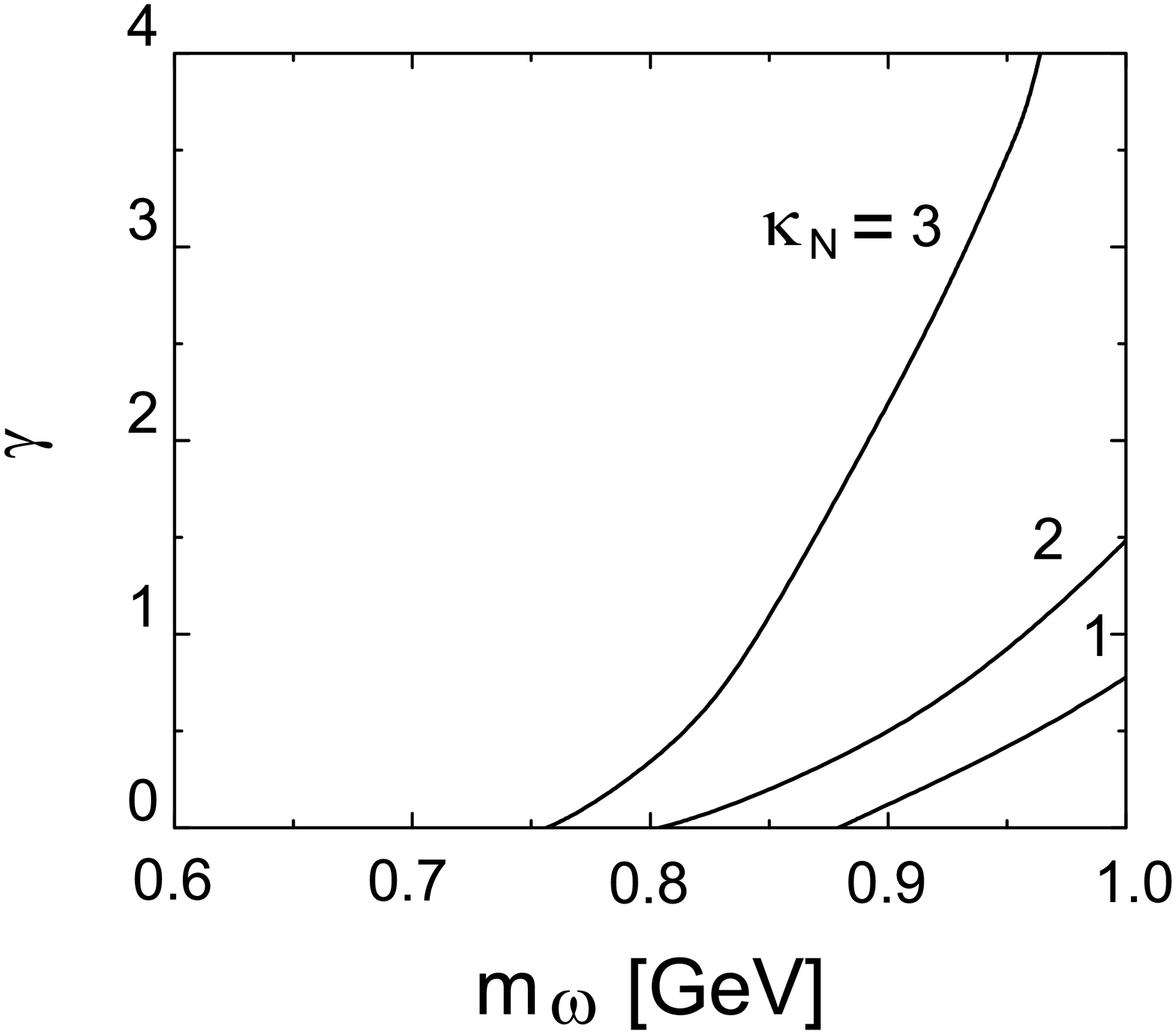,scale=0.33}
\end{minipage}
\begin{minipage}[t]{16.5 cm}
\caption{
As in Fig.~2 but with both $\Re T_{VN}$ 
and $\Im T_{VN}$ from \protect\cite{Lutz}.
\label{fig3}}
\end{minipage}
\end{center}
\end{figure}

To test the importance of this particular form of $T_{VN}$ we proceed to evaluate 
the sum rule\footnote{
At a given baryon density $n$ the continuum threshold $s_V$  
is determined by requiring maximum flatness of 
$m_V (n; M^2,s_V)$ as a function of $M^2$ within 
the Borel window $M_{\rm min}^2\; \cdots \; M_{\rm max}^2$. The minimum Borel 
parameter $M_{\rm min}^2$ is determined such that the terms of order ${\cal O} (M^{-6})$
on the OPE side eq.~(\ref{eq_20}) contribute not more that 10\%.
Selecting such sufficiently large values of  $M_{\rm min}^2$ suppresses higher-order
contributions in the OPE eq.~(\ref{eq_20}) and justifies the truncation.
Typically, $M_{\rm min}^2(10\%)$ is in the order of 0.6 GeV${}^2$.
The values for $M_{\rm max}^2$ are roughly determined by the ''50\% rule'', i.e., 
the continuum part of the hadronic side must not contribute
more than 50\% to the total hadronic side to be sufficiently sensitive
to the resonance part.
According to our experience \cite{Zschocke1,Zschocke2,Zschocke3}, 
$m_V$ is not 
very sensitive to variations of $M_{\rm max}^2$. We can, therefore,  
fix the maximum Borel parameter by $M_{\rm max}^2 = 1.5 \,(2.4) \,{\rm GeV}^2$
for the $\omega$ ($\rho$) meson, in good agreement with the ''50\% rule''. 
The sensitivity of the results on these choices of the Borel window
is discussed in \protect\cite{Zschocke3}.
Flat curves $m_V (n; M^2, s_V)$ within the Borel window
represent a prerequisite for the stability of the sum rule analysis.}
in three steps: 
(i) first neglect $T_{VN}$ at all,
(ii) include only ${\rm Im} T_{VN}$, and
(iii) include both ${\rm Im} T_{VN}$ and ${\rm Re} T_{VN}$ as well.
The results are exhibited in Figs.~1 - 3.
Compared to vacuum, the in-medium changes of the r.h.s. of eq.~(\ref{eq_70})
require for the $\rho$ meson more strength at low energy which is accomplished,
for the given parameterization, by a down-shift of the peak position 
or a larger width. With increasing $\gamma$ the masses are shifted
to larger values. This can be understood in the following way: A larger width
gives more contribution to the integrals on the l.h.s.\ of eq.~(\ref{eq_70})
at lower energy;
this is compensated by an up-shift of the peak position.
Note that $\gamma = 0$ reproduces the zero-width approximation.
 
One observes in Figs.~1 - 3 a strong sensitivity on the density dependence
of the 4-quark condensate. The sensitivity against variations of $\kappa_N$
is larger than that of $\gamma$. (The displayed range of $\gamma$ covers an
enormous range of the widths.)
Remarkable is the tendency of a down-shift
of the $\rho$ mass and a related up-shift of the $\omega$ mass when restricting
the width parameters $\gamma_V$ to smaller values.
The overall pattern seen in figs.~1 - 3 is quite robust. The numerical details,
however, change under variations of $T_{VN}$. To have a fix point let us 
consider $\kappa_N = 3$ (which will later turn out as relevant value).
The peak position of the $\rho$ meson is stable with respect to variations
of $T_{VN}$. This may be attributed to the large vacuum width of $\rho$;
changes of the amount of $T_{VN}$ by $\gamma$ cause only small changes of the
$\rho$ peak position. The $\omega$ meson, in contrast, seems to sit at some
borderline: the variations of $T_{VN}$ can cause an up-shift or a down-shift.
Inclusion of $\Im T_{VN}$ alone results in a tiny effect, while $\Re T_{VN}$
pushes the peak further up. Larger values of $\kappa_N$ are required
to get the $\omega$ meson's peak position down-shifted.  

In contrast to the universal scaling hypothesis \cite{BR_scaling} 
there is no unique in-medium behavior of $\rho$ and $\omega$ mesons.
Rather, the strikingly different values of the subtraction constants
$\Pi_V (0, n)$ cause the different behavior of $\rho$ and $\omega$ mesons,
as stressed in \cite{Hofmann}.\footnote{The detailed analysis of further
terms in the OPE side of the sum rule will be relegated to a separate study.}

\section{$\rho$ -- $\omega$ mass splitting} %%%%%%%%%%%%%%%%%%%%%%%%%

Inspecting Figs.~1 - 3 seems to point to a lacking predictive power of the QSR.
However, if one is interested in the peak position of the
spectral function (usually called ''in-medium mass''), and not in the
very details of the shape of $S^{(V)}$, 
and if one assumes that the essential features of the in-medium
broadening are sufficiently accurate described by $\Im T_{VN}$, 
one has to put $\gamma_V = 1$.
Then one arrives at the in-medium mass splitting of $\rho$ and $\omega$
mesons as displayed in Fig.~4. The mass splitting is surprisingly large
already at normal nuclear density $n_0$. The magnitude of the mass splitting
can be traced back to the particular form of $\Im T_{VN}$ from \cite{Lutz}.
(Using $\Im T_{VN}$ from \cite{Klingl_Weise} results in a smaller splitting.)
The driving force of the $\rho$ -- $\omega$ mass splitting, however, is the
difference of $\Pi^\omega (0,n)$ and $\Pi^\rho (0,n)$. 
 
The $\rho$ and $\omega$ mass shifts are determined
by the still unconstraint parameter $\kappa_N$. Once one of the vector mesons's
peak position is experimentally determined the other one is fixed by
the correlation displayed in Fig.~4, up to some uncertainty caused
by the actual $\Im T_{VN}$ or the freedom in $\gamma_V$.

A presently running experiment \cite{Metag} measures the reaction
$\gamma + A \to X + \omega$ with identifying the $\omega$ via the
$3 \gamma$ Dalitz decay. First data analysis \cite{Metag} seems
to exclude any up-shift, and even a weak down-shift 
of $\omega$ strength
is compatible with the data. If this result gets confirmed
it would require, within the present approach, that $\kappa_N > 3$,
i.e., a strong density dependence of the 4-quark condensate.
In other words, this would give a first direct evidence for a change
of a condensate. Even a null effect for $\omega$ mesons decaying
inside the target nucleus would require a strong density dependence
of the 4-quark condensate, as evidenced by Fig.~4. In fact, at $n_0$ and for 
$\kappa_N = 3$ the 4-quark condensate drops to 60\% of its vacuum value.

\begin{figure}[t]
\begin{center}
\begin{minipage}[t]{16 cm}
\epsfig{file=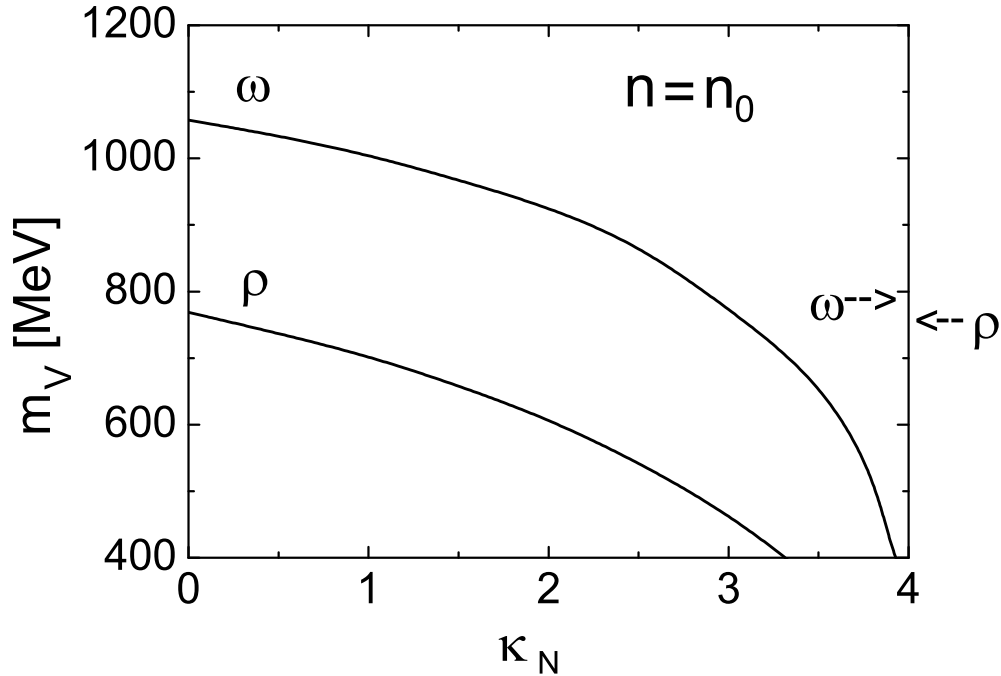,scale=0.5}
\end{minipage}
\begin{minipage}[t]{16.5 cm}
\caption{The peak positions of the $\omega$ (upper curve) and
$\rho$ meson (lower curve) as a function of the parameter $\kappa_N$.
Same choice of parameters as in Fig.~2 for $\gamma = 1$.
The arrows depict the vacuum masses.
\label{fig4}}
\end{minipage}
\end{center}
\end{figure}

\section{Conclusion} %%%%%%%%%%%%%%%%%%%%%%%%%%%%%%%%%%%%%%%%%%%%%%%%

In summary, we present here an analysis of the QCD sum rule for
the in-medium behavior of $\rho$ and $\omega$ mesons. Truncating
the ''QCD side'' of the sum rule at mass dimension 6 and twist 2
we find a strong sensitivity of the $\rho$ and $\omega$ 
peak positions on the density
dependence of the 4-quark condensate. Assuming that we have at our disposal
a sufficiently realistic description of the collision broadening,
described in low-density approximation by the off-shell forward meson-nucleon
scattering amplitude $\Im T_{VN}$, the in-medium mass shifts of $\rho$ and $\omega$
mesons are related essentially to one parameter. Experiments
identifying the $\omega$ decay in nuclear matter then constrain
this parameter, thus fixing also the in-medium $\rho$ mass,
up to uncertainties inherent in the sum rule approach.
If an up-shift of $\omega$ strength can be experimentally excluded,
the present QCD sum rule analysis points to a strong reduction
of the 4-quark condensate already at nuclear saturation density.

On a quantitative level there is some uncertainty caused by the
actual form of $\Im T_{VN}$, in particular for the $\rho$ meson
which influences the $\omega - \rho$ mass splitting. Poorly known
twist-4 condensates modify further this splitting and deserve
additional investigations. Given the importance of the 4-quark
condensate, one could be afraid on the influence of higher
order condensates. Here, the hope is that an appropriate Borel
window suppresses these higher orders. 
We also found some quantitative changes when including explicitly
$\Re T_{VN}$ in the spectral function. But the overall pattern
of the in-medium modifications of $\rho$ and $\omega$ mesons
is stable.

In contrast to $\rho$ and $\omega$ mesons, which are insensitive
against changes of the genuine chiral condensate, $\langle \bar q q \rangle$,
the $\phi$ meson depends sensitively on the chiral strange condensate, 
$\langle \bar s s \rangle$,
and only very weakly on the 4-quark condensate. 

With respect to the future accelerator project SIS200/300 at GSI
an extension of the present approach to the in-medium behavior
of $D$ mesons is challenging.\\[3mm]
{\small 
One of the authors (B.K.) thanks Amand Faessler for inviting him
to the International School of Nuclear Physics, Erice (It), Sep.\ 16 - 24, 2003, 
where the material of this article has been presented.
Many useful discussions with S. Leupold, M. Lutz, V. Metag, U. Mosel,
and Gy. Wolf are gratefully acknowledged. 
The work is supported by BMBF 06DR121 and GSI.}

\end{document}